\documentclass[prb,twocolumn,showpacs,superscriptaddress,floatfix]{revtex4-1}
\usepackage{graphicx,amsfonts,amssymb,amsmath,xspace,dsfont}
\usepackage[colorlinks=true,citecolor=green,linkcolor=red,urlcolor=blue]{hyperref}

\usepackage{color}
\definecolor{g}{rgb}{.1,0.4,.1} 
\definecolor{b}{rgb}{0,0.2,1}
\definecolor{rouge}{rgb}{0.82,0.,0.}
\definecolor{vert}{rgb}{0.,0.82,0.}
\definecolor{orange}{rgb}{1,0.5,0.}
\definecolor{bleu}{rgb}{0.,0.,0.82}
\definecolor{m}{rgb}{0.82,0.,0.82}
\definecolor{vert2}{rgb}{0.,0.5,0.}
\definecolor{rougeclair}{rgb}{1.0,0.7,0.7}

\begin{document}

\title{Russian doll spectrum in a non-Abelian string-net ladder}

\author{Marc Daniel Schulz}
\email{mdschulz@umn.edu}
\affiliation{Department of Physics and Astronomy, University of Minnesota, Minneapolis, Minnesota 55455, USA}
\author{S\'{e}bastien Dusuel}
\email{sdusuel@gmail.com}
\affiliation{Lyc\'ee Saint-Louis, 44 Boulevard Saint-Michel, 75006 Paris, France}
\author{Julien Vidal}
\email{vidal@lptmc.jussieu.fr}
\affiliation{Laboratoire de Physique Th\'eorique de la Mati\`ere Condens\'ee,
CNRS UMR 7600, Universit\'e Pierre et Marie Curie, 4 Place Jussieu, 75252
Paris Cedex 05, France}

\begin{abstract}
We study a string-net ladder in the presence of a string tension. Focusing on the simplest non-Abelian anyon theory with a quantum dimension larger than two, we determine the phase diagram and find a Russian doll spectrum featuring size-independent energy levels as well as highly degenerate zero-energy eigenstates. At the self-dual points, we compute the gap exactly by using a mapping onto the Temperley-Lieb chain. These results are in stark constrast with the ones obtained for Fibonacci or Ising theories.
\end{abstract}


\maketitle
%
%
\section{Introduction}
%
%

Topologically ordered quantum phases of matter have triggered much attention over the last decades in different domains (see Ref.~\onlinecite{Wen13} for a review). In two dimensions, these phases host exotic excitations known as anyons \cite{Leinaas77} whose experimental observation remains one of the important challenges in condensed matter physics. 
Anyons are one of the key ingredients of topological quantum computation \cite{Kitaev03,Preskill_HP,Wang_book} and, in this perspective, it is crucial to understand their behavior as well as transitions they may induce \cite{Bais09_1,Bais09_2,Gils09_2,Gils13,Finch13,Fendley13,Fuchs13,Kong14,Eliens14,Finch14}. To address these issues, many microscopic models have been proposed, among them the string-net model \cite{Levin05} that allows one to generate a wide class of topological phases \cite{Liu14,Lin14} and to study their \mbox{stability \cite{Gils09_1,Gils09_3,Ardonne11,Burnell11_2,Burnell12,Schulz13,Schulz14}}. 

\begin{figure}[t]
\includegraphics[width=0.5\columnwidth]{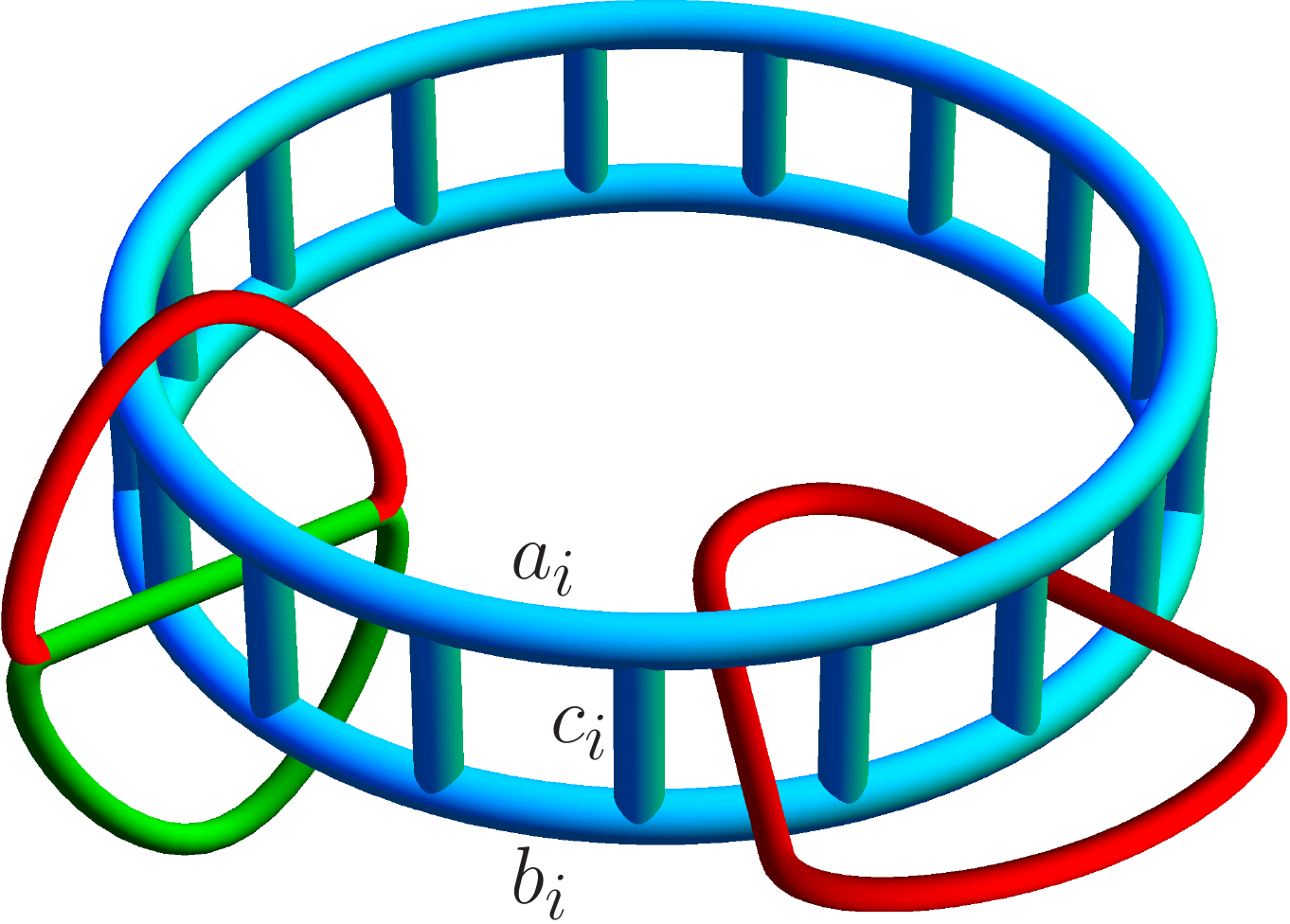}
\caption{(Color online). Ladder with periodic conditions. Lines piercing plaquettes represent flux excitations of the string-net Hamiltonian. Different colors represent different anyons (the trivial label $0$ is invisible by convention). 
Left: a single green-flux excitation with $q_a={\rm red}$ and $q_b={\rm green}$. Right: a two red-flux excitation with $q_a=q_b=0$. 
}
\label{fig:ladder}
\end{figure}

This model can be defined on any trivalent graph \cite{Levin05} so that the simplest system that can be studied is the two-leg ladder displayed in Fig.~\ref{fig:ladder}. Strictly speaking, there is no topological order in gapped one-dimensional systems (due to absence of long-range entanglement \cite{Hastings07,Chen10_1}). However, some mechanisms driving phase transitions in the so-called string-net ladder are also relevant in two dimensions \cite{Gils09_1}. 
So far, the string-net ladder has first been investigated for Fibonacci \cite{Gils09_1} and Ising \cite{Gils09_3} anyons  (see also Ref.~\onlinecite{Ardonne11} for the non-unitary Yang-Lee theory). Surprisingly enough, in the presence of a string tension, all these non-Abelian theories display a qualitatively similar phase diagram consisting of two gapped phases separated by a critical point and a critical phase. Nevertheless, one should keep in mind that all these theories contain only one non-Abelian anyon. 

In this work, we study the string-net model in the ladder geometry in the presence of a string tension for the simplest theory with two non-Abelian anyons and we show that it gives rise to properties absent in the above mentioned theories. In particular, the critical phase and the critical point observed for Fibonacci and Ising anyons are found to be gapped for this theory. Moreover, we find a rich Russian doll spectrum with some size-independent energy levels and highly degenerate zero-energy eigenstates. This paper is organized as follows. In Sec.~\ref{sec:model}, we introduce the string-net model with a string tension and we analyze ground-state degeneracies in simple limiting cases. We also discuss conserved quantities that are crucial to analyze the spectrum. These results are valid for any anyon model. In Sec.~\ref{sec:theory}, we focus on a theory which has two non-Abelian particles, the so-called $(A_1,5)_{1/2}$ theory, and we derive the energy spectrum of the string-net model without string tension. In Sec.~\ref{sec:PD}, we study numerically the phase diagram for this theory as a function of the string tension. This diagram consists in three gapped phases separated by first-order transitions. 
In Sec.~\ref{sec:self-dual}, we analyze the self-dual points and we compute the low-energy gap analytically in the thermodynamical limit by using a mapping onto the $XXZ$ chain. We conjecture that such a mapping holds for any anyon theory. This conjecture allows us to conclude that self-dual points are only gapless when the total quantum dimension is lower than two. 
%
%
%
%

%
%
\section{Model and symmetries}
\label{sec:model}
%
%
In the string-net model, microscopic degrees of freedom are defined on edges of a trivalent graph. For a theory with $N$ anyons, each edge can be in $N$ different states. The Hilbert space ${\mathcal H}$ is then defined by the set of states compatible with the fusion rules at each vertex of the graph \cite{Levin05}. For any trivalent graph, the dimension of the Hilbert space can be expressed as a function of the fusion matrices  of the considered theory  \cite{Gils09_1}. 

The string-net Hamiltonian with a string tension is defined as
%
%
\begin{equation}
 \label{eq:ham}
H=- J_{\rm p} \sum_p  \delta_{\Phi(p),0} - J_{\rm r} \sum_r \delta_{l(r),0}.
\end{equation} 
%
%
The first term is  the string-net Hamiltonian introduced by Levin and Wen \cite{Levin05} defined on a ladder. It involves the projector $\delta_{\Phi(p),0}$ onto states with no flux $\Phi(p)$ through \mbox{plaquette $p$}. Explicit form of this operator in the edge basis is given in Ref.~\onlinecite{Gils09_1}. The second term is diagonal in the link (edge) basis since $\delta_{l(r),0}$ is the projector onto the state $0$, acting on the rung $r$. Without loss of generality, we set $J_{\rm p}=\cos \theta$ and $J_{\rm r}=\sin \theta$.

For any modular theory, $H$  has a remarkable property first identified in Ref.~\onlinecite{Gils09_1}. Indeed, for twisted boundary conditions, its spectrum is exactly self-dual, i.e., it is invariant under the transformation $\theta \leftrightarrow \pi/2 -\theta$ that amounts to exchange $J_{\rm p}$ and $J_{\rm r}$. In particular, energy spectra for $\theta= \pi/2$ and $3\pi/2$ are the same as for \mbox{$\theta= 0$} and $\pi$, respectively. 
For periodic boundary conditions, the self-duality is broken for finite-size systems but it is restored in the thermodynamical limit up to degeneracies. This discrepancy can readily be observed on the ground-state degeneracy which,  for any modular theory with $N$ anyons and for \mbox{$\theta=\pi/2$}, is  $N$-fold ($N^2$-fold) degenerate for twisted (periodic) boundary conditions. By contrast, for $\theta=0$, the ground state is $N$-fold degenerate for twisted and for periodic boundary conditions.

For periodic boundary conditions considered in the following, $H$ commutes with the translation operator along the ladder direction $T_x$ and also with the reflection operator in the transverse direction $T_y$. 
Interestingly, $H$ also commutes with less obvious operators that measure the flux above and below the ladder. Indeed, a simple picture of the string-net model consists in representing each excitation by a flux-line piercing a plaquette (see Fig.~\ref{fig:ladder}). These flux lines are endless (no monopoles) so that they have to be connected according to the fusion rules. This physical representation allows one to compute level degeneracies for $\theta=0$ [see Eq.~(\ref{eq:deg})] but it also allows to identify two additional conserved quantities for any $\theta$. To define the corresponding operators, it is useful to define the projectors onto flux $\alpha$ above $(a)$ and below $(b)$ the ladder
%
%
\begin{eqnarray}
\label{eq:proj}
P^{(\alpha)}_a |a,b,c \rangle&=& S_{1 \alpha} \sum_\beta S_{\alpha  \beta} \sum_{a'} \prod_{i=1}^{N_{\rm p}}\Big[F_{a'_{i+1}}^{c_i a_i \beta}\Big]_{{a_{i+1}}{a'_i}} \hspace{-3.5mm} |a',b,c \rangle,  \quad \\
P^{(\alpha)}_b |a,b,c \rangle&=& S_{1 \alpha}\sum_\beta S_{\alpha  \beta} \sum_{b'} \prod_{i=1}^{N_{\rm p}} \Big[F_{b'_{i+1}}^{c_i b_i \beta}\Big]_{{b_{i+1}}{b'_i}} 
\hspace{-3.5mm}|a,b',c \rangle, 
\end{eqnarray} 
%
%
where $N_{\rm p}$ is the number of plaquette and $|a,b,c \rangle$ is a state defined by the set of labels $\{a_i,b_i,c_i\}_{i=1,\cdots,N_{\rm p}}$ on all edges (see  Fig.~\ref{fig:ladder}). The $S$-matrix and $F$-symbols are given by the theory of interest (see, e.g., Ref.~\onlinecite{Rowell09}). From these two sets of (orthogonal) projectors, one can define two operators measuring the flux above and below the ladder as  $Q_i=\sum_\alpha \alpha \:P^{(\alpha)}_i$  with eigenvalues \mbox{$q_i=\alpha$} (here, \mbox{$i=a,b$}). There are two reasons for which $P^{(\alpha)}_i$, and hence $Q_i$, commutes with $H$: (i) by construction, any closed-string operator (in the canonical link basis) commutes with the string-net Hamiltonian \cite{Levin05} and (ii) $P^{(\alpha)}_i$ does not change the rung degrees of freedom ${c_i}$ so that it commutes with the second term in $H$. However, since these fluxes can be different above and below the ladder (see Fig.~\ref{fig:ladder} for a concrete example), the corresponding operators do not commute with $T_y$. Thus, there are three conserved quantities: the momentum along the chain direction and fluxes  $(q_a, q_b)$. Note that another choice has been proposed in \mbox{Refs.~\onlinecite{Gils09_1,Gils09_3,Ardonne11}}.

%
%
\section{The $(A_1,5)_{1/2}$ theory}
\label{sec:theory}
%
%
In the following, we focus on the simplest theory with more than one non-Abelian particle. This theory known as  $(A_1,5)_{1/2}$  contains three anyons  $0, 1$, and $2$, obeying SU$(2)_5$ fusion rules restricted to integer labels (see Ref. \onlinecite{Rowell09} and Appendix \ref{app:data} for more details)
%
%
\begin{eqnarray}
0 \times a &=& a \times 0= a, \:\: \forall a \in \{0,1,2\}, \label{eq:fusion1} \\
1 \times 1 &=& 0+1+2, \label{eq:fusion2} \\
1 \times 2 &=& 2 \times 1=1+2,  \label{eq:fusion3}\\
2 \times 2 &=& 0+1. \label{eq:fusion4}
\end{eqnarray} 
%
%
Setting $d = 2\cos(\pi/7)$, quantum dimensions of these anyons are $(d_0, d_1, d_2) = (1, d^2-1, d)$ so that the total quantum dimension is $D=\sqrt{\sum_i d_i^2}=\frac{\sqrt{7}}{2 \sin(\pi/7)}$.  
For a ladder with $N_{\mathrm p}$ plaquettes, the Hilbert space dimension is given by
%
%
\begin{eqnarray}
 \label{eq:dimH}
 \dim \mathcal{H}&=&\left(-3 d^2+2 d+9\right)^{N_{\rm p}}+\left(d^2-3d+4\right)^{N_{\rm p}} \nonumber \\
&&+\left(2 d^2+d+1\right)^{N_{\rm p}},
\end{eqnarray} 
%
%
which grows as  $D^{2 N_{\rm p}}$ in the thermodynamical limit.

By construction, for $\theta=0$,  the system is in a doubled $(A_1,5)_{1/2}$ phase denoted by D$(A_1,5)_{1/2}$ in the following. As such, the degeneracy of the eigenstates depends on the graph topology. In the ladder, the degeneracy of the $k$th energy level \mbox{$E_k=-N_{\rm p}+k$}   is  \cite{Dusuel14}
%
%
\begin{eqnarray}
\label{eq:deg}
\mathcal{D}_k&=&\left(
\begin{array}{c}
N_{\mathrm p}
\\
k
\end{array}
\right)\times \\
&&
\left[\left(-3 d^2+2 d+8\right)^k+\left(d^2-3 d+3\right)^k+(2 d^2+d)^k\right], \nonumber
\end{eqnarray} 
%
%
where the binomial coefficient stems from the different ways to choose $k$ plaquettes carrying the flux excitations among $N_{\mathrm p}$. In particular, one finds $\mathcal{D}_0=3$ ground states and one can  check that $\displaystyle{\dim \mathcal{H}=\sum_{k=0}^{N_{\rm p}} \mathcal{D}_k}$.

For $\theta=\pi$, the ground-state degeneracy  is \mbox{$\mathcal{D}_{N_{\rm p}}\simeq (D^2-1)^{N_{\mathrm p}}$} and the energy of the $k$th energy level is given by \mbox{$E_k=k$}.

%
%
\begin{figure}[t]
\includegraphics[width=0.9\columnwidth]{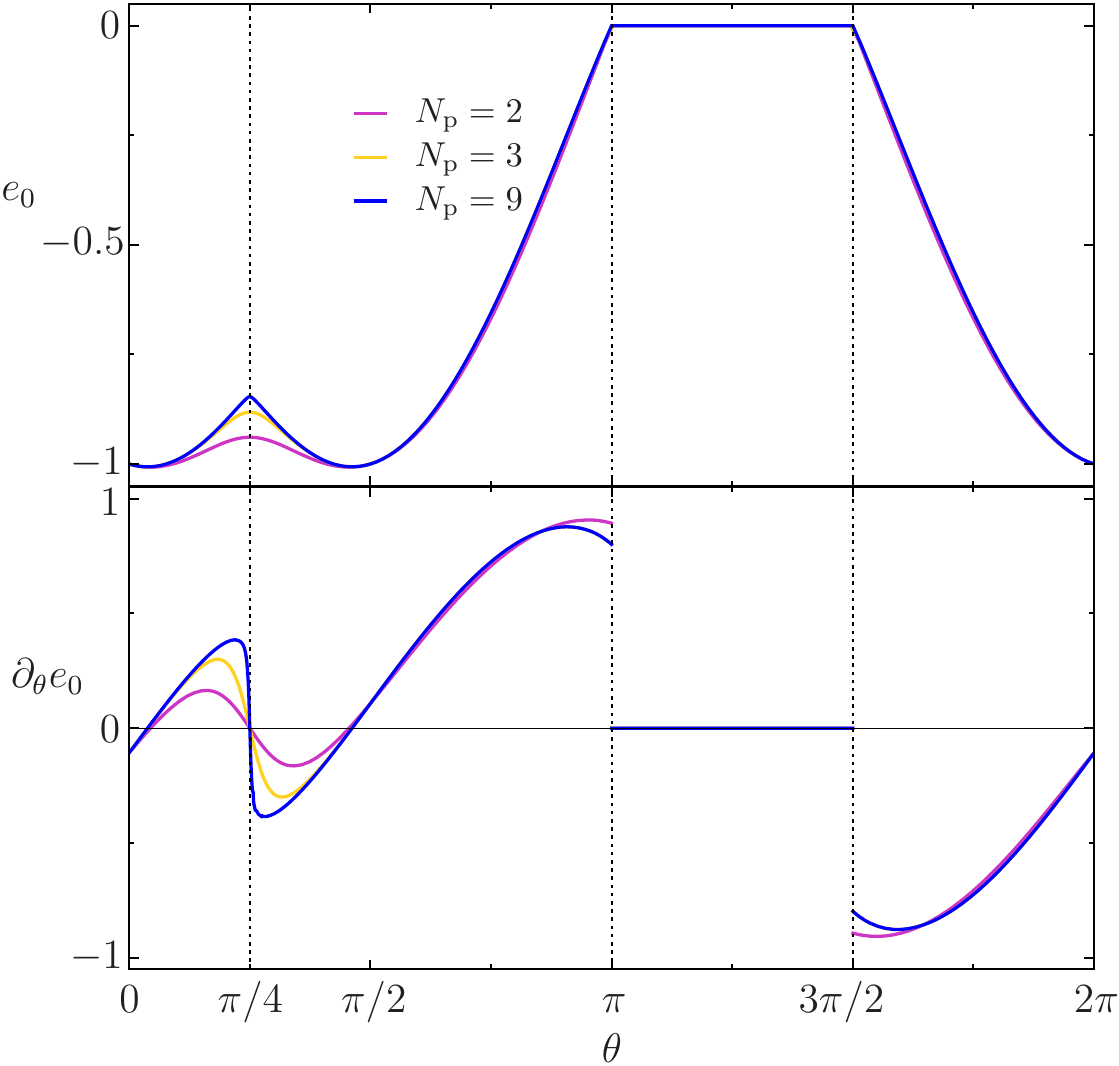}
\caption{(Color online). Top: ground-state energy per plaquette $e_0$ as a function of $\theta$ for various system sizes. Bottom: first derivative of $e_0$ with respect to $\theta$.}
\label{fig:gse}
\end{figure}
%
%

%
%
\section{Phase diagram}
\label{sec:PD}
%
%

\subsection{Generalities}
\begin{figure}[t]
\includegraphics[width=0.5\columnwidth]{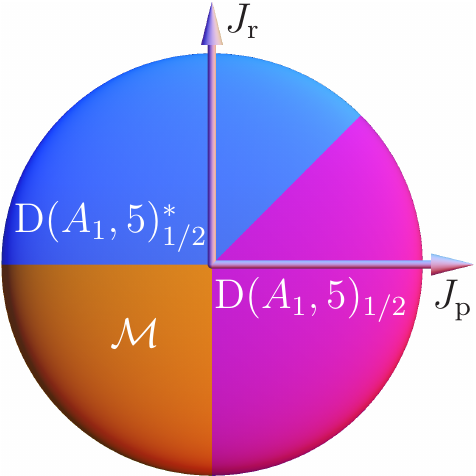}
\caption{(Color online). Phase diagram consisting of three gapped phases separated by first-order transitions at \mbox{$\theta=\pi/4, \pi$, and $3\pi/2$}. Self-duality manifests itself as a symmetry with respect to the line $J_{\mathrm p}=J_{\mathrm r}$.}
\label{fig:diagram}
\end{figure}

For arbitrary $\theta$, $H$ is not exactly solvable and it is useful to gain some insights from numerics. 
To study the phase diagram, we performed exact diagonalizations (ED) of the Hamiltonian for system sizes up to  \mbox{$N_{\rm p}=9$} for which  \mbox{$\dim \mathcal{H} \simeq 5 \cdot 10^8$}. The ground-state energy per plaquette $e_0$ and its derivative with respect to $\theta$ for various system sizes is displayed in Fig.~\ref{fig:gse}. As can be seen,  even for finite-size systems, $\partial_\theta e_0$ is discontinuous for $\theta=\pi$  and $3\pi/2$ indicating a level crossing and hence first-order transitions at these points. 
For $\theta=\pi/4$, one clearly sees the formation of a singularity but one cannot infer its nature from numerics. However, as discussed in the next section, we will see that this is also a first-order transition point.
Thus, the phase diagram consists in three phases separated by first-order transitions located at $\theta=\pi/4,\pi$, and $3\pi/2$ (see Fig.~\ref{fig:diagram}). 

%
%
\subsection{The D$(A_1,5)_{1/2}$ phase~: $\theta \in [3\pi/2,\pi/4]$}
%
%

For \mbox{$\theta \in [3\pi/2,\pi/4]$}, the system is in the D$(A_1,5)_{1/2}$ phase which is adiabatically connected to the point $\theta=0$ and hence gapped (high-order perturbative series expansions  of the gap and of the ground-state energy in the  limit $|J_{\mathrm r}| \ll J_{\mathrm p}$ can be found in Appendix \ref{app:series}).

Since  the spectrum is self-dual, a gapped dual phase D$(A_1,5)_{1/2}^*$ is obtained for \mbox{$\theta \in [\pi/4,\pi]$}. Similar gapped phases are found exactly in the same range for Fibonacci and Ising theories.

 %
%
\subsection{The $\mathcal{M}$ phase~: $\theta \in [\pi,3\pi/2]$}
%
%

The most intriguing result is the existence of a gapped phase (denoted by $\mathcal{M}$ in the following) for \mbox{$\theta \in [\pi,3\pi/2]$} instead of critical phases found in this range for previously studied theories \cite{Gils09_1,Gils09_3,Ardonne11}.
%
%
%
%
Unfortunately, the huge ground-state degeneracy at $\theta=\pi$ and $3\pi/2$ prevents to perform a perturbative analysis of the low-energy spectrum in this parameter range. However, as we shall now see, ED results unveil two striking properties.

\begin{figure}[t]
\includegraphics[width=.92\columnwidth]{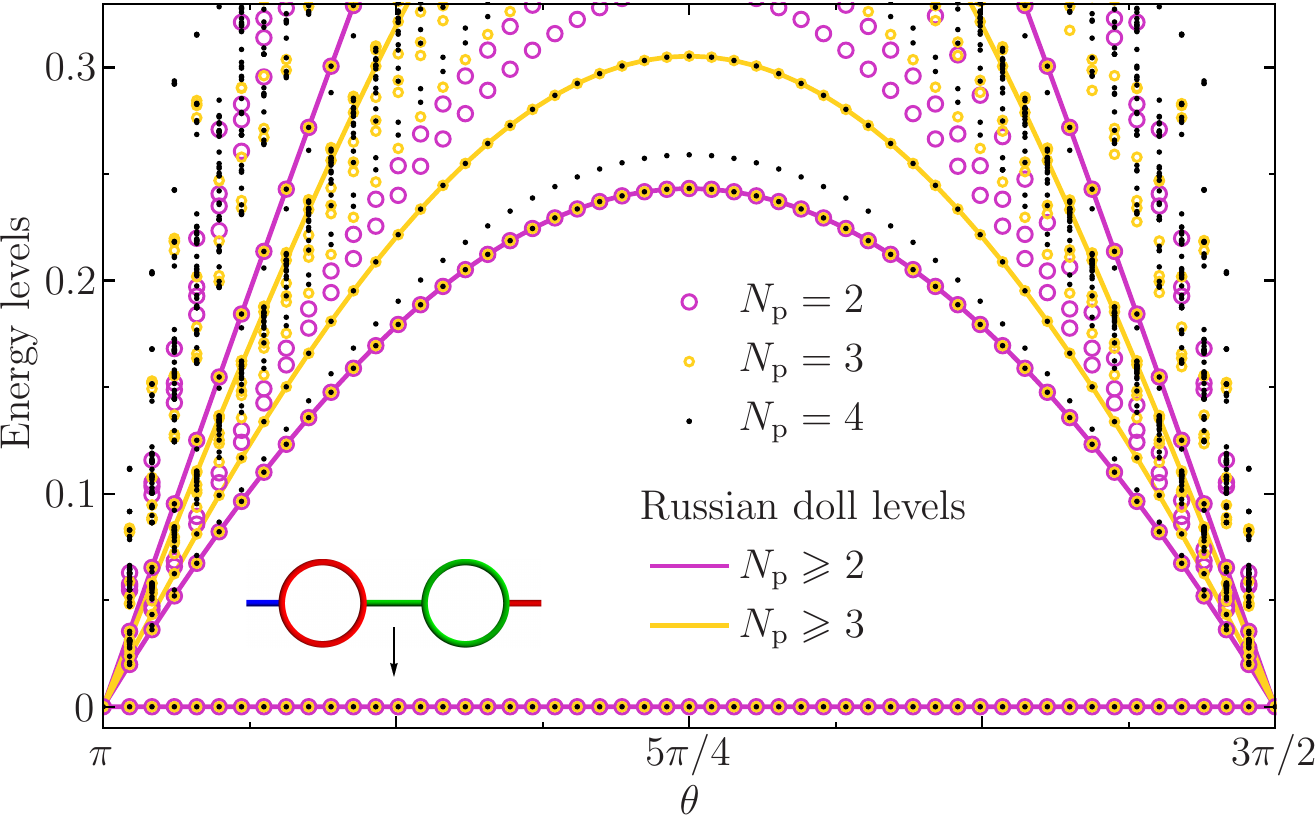}
\caption{(Color online). Low-energy spectrum in the  $\mathcal{M}$ phase. Size-independent levels are highlighted. The gap is given by Eq.~(\ref{eq:gap_exact}). Inset: a possible building block for zero-energy eigenstates described in the bubble basis \cite{Gils09_1} (red=2, green=1, and blue=0).  }
\label{fig:russian_doll}
\end{figure}

 (i) The first one concerns the ground-state energy which equals zero in the whole range $\theta \in [\pi,3\pi/2]$ (see Fig.~\ref{fig:gse}). Despite the fact that \mbox{$[ \delta_{l(r),0},\delta_{\Phi(p),0} ] \neq 0$} if the rung $r$ belongs to the plaquette $p$, it is indeed possible to build zero-energy eigenstates of these two non-commuting projectors. Such states are eigenstates for any $\theta$ but, interestingly, they become ground states for $\theta \in [\pi,3\pi/2]$. A simple example of such a state can be easily sketched by switching from the ladder basis to the bubble basis (see  Supplementary information in Ref.~\onlinecite{Gils09_1} for details about this basis transformation). In this representation, a bubble with different inner and outer legs is a zero-energy eigenstate of the corresponding plaquette operator $\delta_{\Phi(p),0}$ whereas two neighboring bubbles of different type ensure a zero-energy eigenstate of the (shared-) link operator $\delta_{l(r),0}$. For the $(A_1,5)_{1/2}$ theory, these two constraints can easily be satisfied simultaneously (see Fig.~\ref{fig:russian_doll}). Nevertheless, we checked that such a construction does not exhaust all possible zero-energy states whose degeneracy is numerically found to grow exponentially with the system size.
As a side remark, let us mention that we also find zero-energy eigenstates (stemming from a different construction) in the Ising ladder but with a degeneracy only growing linearly with the system size. Furthermore, such states are absent for Fibonacci anyons.  

 (ii) The second unusual property is related to the low-energy gap $\Delta$ which is numerically found to be dependent of $\theta$ but surprisingly independent of the system size (see Fig.~\ref{fig:russian_doll}). 
Although we have not been able to build the corresponding states for any sizes, we computed $\Delta$ exactly by solving the $N_{\mathrm p}=2$ problem and we get
%
%
\begin{equation}
\label{eq:gap_exact}
\Delta_\theta=-\frac{1}{2} \left[\cos \theta +\sin \theta+\sqrt{1+2\left(\frac{8}{D^2}-1\right)\cos \theta \sin \theta} \right].
\end{equation} 
%
%
The corresponding eigenstates are all translation invariant and their degeneracy is numerically found to grow exponentially with the system size. Unfortunately, contrary to zero-energy states discussed above, we have not been able to find a simple procedure to build any of these low-energy states. 

This size-independent gap suggests that, for any \mbox{$\theta\in[0,2\pi]$}, there might exist other size-independent energy levels and this is indeed the case.  For instance, we find that levels with energies $-\cos \theta$ and $-\sin \theta$ exist for all $N_{\mathrm p}$.
This Russian doll  ($\mathcal M$atrioshka) spectrum is an intriguing result that clearly deserves further investigations and a deeper understanding especially since such a structure does not occur for Fibonacci and Ising theories.
%
%
%
%

%
%
\section{Self-dual points}
\label{sec:self-dual}
%
%
As mentioned in the previous section, $H$ is not exactly solvable for arbitrary $\theta$. However, it turns out that, apart from the trivial cases $\theta=0 \mod \pi/2$, there are two additional points for which $H$ is exactly solvable. 

For \mbox{$\theta=\pi/4$}, the Hamiltonian can be written as \mbox{$H_{\pi/4}=-\frac{\sqrt{2}}{D} \frac{1}{2}\sum_i e_i$}  where $e_i$ operators form a representation of  the Temperley-Lieb algebra  
%
%
\begin{equation}
e_i^2=D \; e_i, \:\: e_i e_{i+1}e_i=e_i, \:\: [e_i,e_j]=0 \:\: {\rm for} \:\: |i-j|\geqslant 2.
\end{equation} 
%
%
Derivation of this result is the same as for Fibonacci and Ising theories \cite{Gils09_1,Gils09_3} so that we do not reproduce it here. Unfortunately, we are not able to prove this mapping for arbitrary modular theories but we conjecture that it is always valid. 

Using the correspondence between this Temperley-Lieb chain and the {\em antiferromagnetic} $XXZ$ chain with an anisotropy parameter $D/2$ \cite{Aufgebauer10}, we conclude that the spectrum is gapless for $D \leqslant 2$ and gapped for $D>2$.
Thus, for a given theory,  if \mbox{$\theta=\pi/4$} is a transition point, it is a criti\-cal point for $D \leqslant 2$ and a first-order transition point for $D>2$. 

For the $(A_1,5)_{1/2}$ theory, the total quantum dimension $D$ is strictly larger than two (see Sec.~\ref{sec:theory}) and one can compute the exact value of the gap in the thermodynamical limit \cite{Cloiseaux66}. Setting $\cosh\Phi= D/2$, one gets 
%
%
\begin{equation}
\Delta_{\pi/4} =\frac{\sqrt{2}}{D}\sinh \Phi  \sum_{n=-\infty}^{+\infty} \frac{(-1)^n}{\cosh (n \Phi)} \simeq 0.045237.
\label{eq:gap_XXZ_AF}
\end{equation} 
%
%

\begin{figure}[t]
\includegraphics[width=0.9\columnwidth]{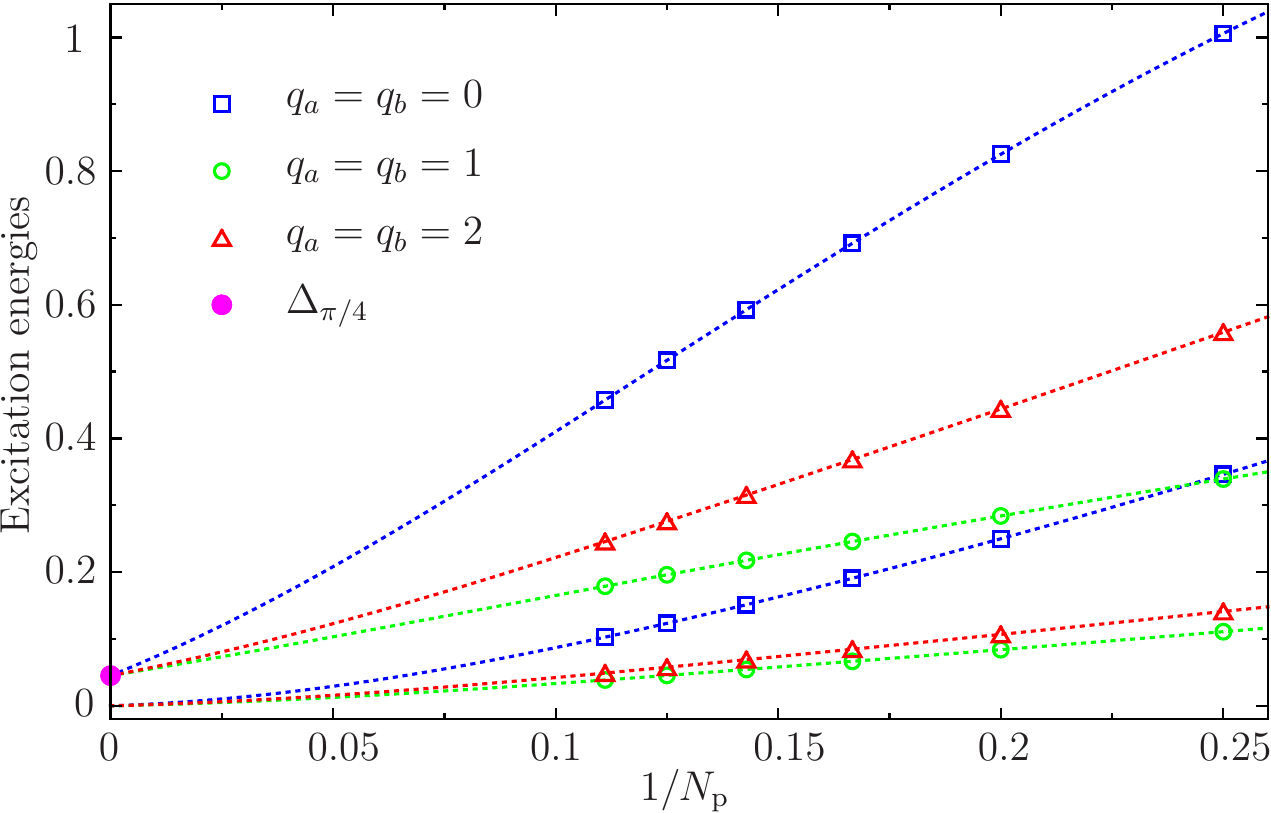}
\caption{(Color online). Finite-size behavior of the two first excitation energies in sectors $q_a=q_b$ for  $\theta=\pi/4$. Dotted lines are guide for the eyes.}
\label{fig:spectrum}
\end{figure}

To extract the gap $\Delta_{\pi/4}$ from ED, one must carefully analyze the spectrum in each topological sector and keep in mind that, with periodic boundary conditions, ground-state topological sectors in the D$(A_1,5)_{1/2}$ phase are such that $q_a=q_b$. By contrast, in the D$(A_1,5)_{1/2}^*$ phase there is one ground state in each sector $(q_a,q_b)$ so that the ground state is expected to be (9+3)-fold degenerate at the transition. In Fig.~\ref{fig:spectrum},  we plot the first excitation energies in sectors $q_a=q_b$. As can be seen, the lowest one goes to zero whereas the second one converges to $\Delta_{\pi/4}$ in the thermodynamical limit.

%
%
%
%

For $\theta=5\pi/4$, one has $H_{5\pi/4}=-H_{\pi/4}$ so that its low-energy properties are now those of the {\em ferromagnetic} $XXZ$ chain \cite{Koma97} which is also only gapped for $D>2$. 
For the $(A_1,5)_{1/2}$ theory, one then gets
%
%
\begin{equation}
\Delta_{5\pi/4}=\frac{\sqrt{2}}{D} \left(\frac{D}{2}-1 \right)\simeq 0.243266,
\label{eq:gap_XXZ_F}
\end{equation} 
%
%
in agreement with Eq.~(\ref{eq:gap_exact}). Note that, for this anyon theory, the point $\theta=5\pi/4$ is not a transition point. 
%
%
\section{Summary and perspectives}
%
%
We studied the phase diagram of a string-net ladder in the presence of a string tension. For the anyon theory considered here, namely the $(A_1,5)_{1/2}$ theory, we found that this diagram consists in three gapped phases separated by first-order transitions. This strongly contrasts with Fibonacci \cite{Gils09_1} and Ising \cite{Gils09_3} theories for which a critical point is found at $\theta=\pi/4$ and a critical phase for $\theta \in [\pi,3\pi/2]$. As explained in Sec.~\ref{sec:self-dual} using a mapping onto the antiferromagnetic $XXZ$ chain, this originates from the fact that these two theories have a total quantum dimension smaller than two. 
For the $(A_1,5)_{1/2}$ theory, we found a Russian doll spectrum which is found neither for the Fibonacci theory nor for the Ising theory. However, we do not know whether such a curiosity exists for other anyon theories. 
We also described a simple way to build  zero-energy eigenstates which can be easily extended to other non-Abelian anyon theories such as the one obeying SU$(2)_{k>2}$ fusion rules. 

To conclude, let us emphasize that the phase diagram has been analyzed by considering the ground state and the first-excited state. However, interesting features also emerge in higher-energy levels. In particular, when the total quantum dimension is larger than two, one can show that elementary excitations in the ground-state sector are anyonic bound states suggesting a possible relationship between the nature of phase transitions and the existence of low-energy bound states.

\acknowledgments

We wish to thank E. Ardonne and P. Finch for  insightful discussions.

\appendix

\section{Data of the $(A_1,5)_{1/2}$ Unitary Modular Tensor Category}
\label{app:data}

Thereafter, we give  all informations about the $(A_1,5)_{1/2}$ unitary modular tensor category (UMTC). Details about their derivation can be found in Ref.~\onlinecite{Rowell09}. The $(A_1,5)_{1/2}$ UMTC is a rank-three modular categories which means that it contains three types of anyons  $0,1,2$ (denoted by $1,\beta,\alpha$, respectively, in Ref.~\onlinecite{Rowell09}). 

%
%
\subsection{Fusion rules}
%
%
These anyons obey the following fusion rules  
%
%
\begin{eqnarray}
0 \times a &=& a \times 0= a, \:\: \forall a \in \{0,1,2\},  \\
1 \times 1 &=& 0+1+2, \\
1 \times 2 &=& 2 \times 1=1+2,  \\
2 \times 2 &=& 0+1.
\end{eqnarray} 
%
%

%
%
\subsection{$S$-matrix}
%
%
The modular $S$-matrix is given by \cite{Rowell09}:
%
%
\begin{eqnarray}
\label{eq:deg_Fibo}
S &=&\frac{1}{D}\left(
\begin{array}{c c c}
d_0& d_1 & d_2 \\
d_1 & -d_2 & d_0 \\
d_2 & d_0& -d_1
\end{array}
\right),
\end{eqnarray} 
%
%
where $(d_0,d_1,d_2)=(1,d^2-1,d)$, $d=2\cos(\pi/7)$ are the quantum dimensions of the anyons, and \mbox{$D= \sqrt{\sum_i d_i^2}=\frac{\sqrt{7}}{2 \sin(\pi/7)}$} is the total quantum dimension. Indices of $S$ are given by the anyon type and thus run here from zero to two. 

%
%
\subsection{$F$-symbols}
%
%
The key ingredient to build the string-net Hamiltonian is the set of $F$-symbols \cite{Levin05,Gils09_1}. These $F$-symbols are defined as follows:
%
%
\begin{align}
\begin{array}{c}\includegraphics[width=1cm]{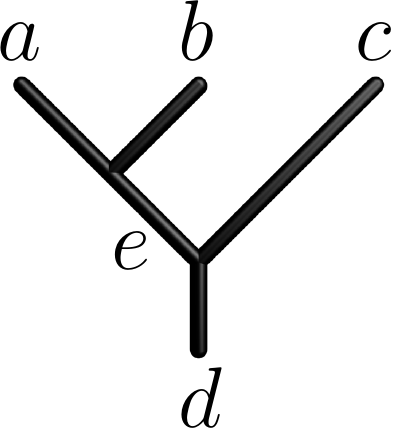}\end{array}
= 
\sum_f \left[F^{abc}_d \right] _{ef} \begin{array}{c}\includegraphics[width=1cm]{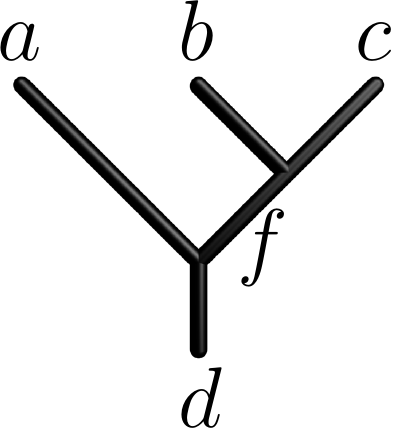}\end{array}.
\label{eq:Fmove}
\end{align}
%
%

General formul{\ae} for any SU$(2)_k$ theory can be found in Ref.~\onlinecite{Kirillov88} but we give them here explicitly for the case $k=5$ which is relevant for our purpose. For $k=5$ and considering only integer labels of SU$(2)_k$, there are several nontrivial  $F-$ symbols, which can be written in the following matrix form

%
%
\begin{eqnarray}
F_1^{1 1 1} =\frac{1}{d_1}\left(
\begin{array}{c c c}
1& -\sqrt{d_1} & \sqrt{d_2} \\
-\sqrt{d_1} & 1/d_1 &  d_2\sqrt{d_2/d_1} \\
\sqrt{d_2} &  d_2\sqrt{d_2/d_1} & d_2/d_1
\end{array}
\right), \qquad 
\end{eqnarray} 
%
%

%
%
\begin{eqnarray}
F_2^{1 1 1} =\frac{1}{d_1}\left(
\begin{array}{c c c}
0&0&0\\
0&-d_2&  \sqrt{d_2} \\
0&\sqrt{d_2} & d_2
\end{array}
\right), \qquad 
\end{eqnarray} 
%
%

%
%
\begin{eqnarray}
F_2^{1 1 2} =\frac{1}{\sqrt{d_1 d_2}}\left(
\begin{array}{c c c}
0&\sqrt{d_1}&  -\sqrt{d_2} \\
0&-\sqrt{d_2} & -\sqrt{d_1} \\
0&0&0
\end{array}
\right), \qquad 
\end{eqnarray} 
%
%

%
%
\begin{eqnarray}
F_2^{1 2 1} =\frac{1}{d_1}\left(
\begin{array}{c c c}
0&0&0 \\
0&1& -\sqrt{d_1 d_2} \\
0&-\sqrt{d_1 d_2} & -1
\end{array}
\right), \qquad 
\end{eqnarray} 
%
%

%
%
\begin{eqnarray}
F_1^{1 2 2} =\frac{1}{\sqrt{d_1 d_2}}\left(
\begin{array}{c c c}
0&0&0 \\
\sqrt{d_1}&  -\sqrt{d_2} &0\\
-\sqrt{d_2} &- \sqrt{d_1}&0
\end{array}
\right), \qquad 
\end{eqnarray} 
%
%

%
%
\begin{eqnarray}
F_2^{2 2 2} =\frac{1}{d_2}\left(
\begin{array}{c c c}
1&  -\sqrt{d_1} &0\\
-\sqrt{d_1} &- 1&0 \\
0&0&0\end{array}
\right).\qquad 
\end{eqnarray} 
%
%
As for the $S$-matrix, indices of these matrices run from zero to two.
Other $F$-symbols are equal to 1 if fusion channels are allowed and 0 otherwise. Missing symbols can be found by using the following identities: 
%
%
\begin{equation}
F_d^{abc}=F_a^{d c b}=F_c^{b a d}=F_b^{c d a}.
\end{equation} 
%
%

%
%
\subsection{$R$-symbols}
%
%
The $R$-symbols are defined as follows:
\begin{align}
\begin{array}{c}\includegraphics[width=1cm]{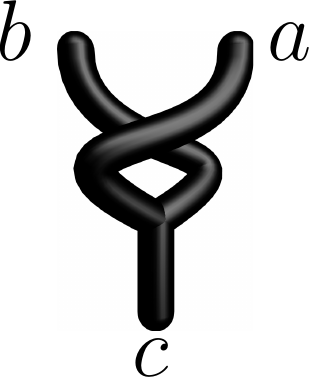}\end{array} = R^{ab}_c \begin{array}{c}\includegraphics[width=1cm]{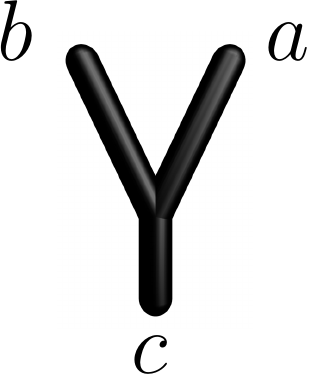}\end{array}.
\label{eq:Rmove}
\end{align}

Note that the $R$-symbols are only non-zero for vertices obeying the fusion rules.
As braiding with the vaccuum is trivial, one has:
%
%
\begin{equation}
R^{0a}_a=R^{a0}_a=1,  \:\: \forall a.
\end{equation} 
%
%
Other $R$-symbols are given by:
\begin{align}
&R^{11}_0 =\mathrm{e}^{-\frac{4\mathrm{i}\pi}{7}} , R^{22}_0 =\mathrm{e}^{\frac{2\mathrm{i}\pi}{7}},\\
&R^{11}_1 =\mathrm{e}^{\frac{5\mathrm{i}\pi}{7}} , R^{12}_1 =\mathrm{e}^{-\frac{6\mathrm{i}\pi}{7}} , 
R^{21}_1 =\mathrm{e}^{-\frac{6\mathrm{i}\pi}{7}} , R^{22}_1 =\mathrm{e}^{-\frac{3\mathrm{i}\pi}{7}}, \\
&R^{11}_2 =\mathrm{e}^{\frac{2\mathrm{i}\pi}{7}} , R^{12}_2 =\mathrm{e}^{\frac{5\mathrm{i}\pi}{7}} , 
R^{21}_2 =\mathrm{e}^{\frac{5\mathrm{i}\pi}{7}}. &
\end{align}

Note that these $R$-symbols are the complex conjugates of the one given in Ref.~\onlinecite{Rowell09} where a typo for $R^{\alpha \alpha}_\beta$ must be corrected. 
%
%
\subsection{Gauge transformations}
%
%

Finally, we would like to mention that this set of $F$-symbols and $R$-symbols can be changed into a new set $\widetilde F$ and $\widetilde R$ according to the following gauge transformation (see, e.g., Ref.~\cite{Bonderson_thesis})
%
%
\begin{equation}
\Big[\widetilde F_d^{abc}\Big]_{ef}=\frac{u^{af}_d u^{bc}_f} {u^{ab}_e u^{ec}_d} \Big[F_d^{abc}\Big]_{ef}, \quad
\widetilde R_c^{ab}=\frac{u^{b a}_c} {u^{a b}_c} R_c^{ab},
\end{equation} 
%
%
where $u$'s are complex numbers. In order to exactly map  our Hamiltonian onto a Temperley-Lieb chain Hamiltonian at the self-dual points $\theta=\pi/4, 5\pi/4$ one needs to choose:
%
%
\begin{equation}
u^{1 1}_2=u^{1 2}_1=u^{2 1}_1=\rm{i},
\end{equation} 
%
%
and $u^{i j}_k=1$ otherwise.

%
%
\section{Series expansions in the limit $|J_{\rm{r}}| \ll J_{\rm{p}}$}
\label{app:series}
%
%

In the following, we give the series expansions for the ground-state energy per plaquette $e_0$ and for the one-quasiparticle gaps $\Delta^+$ and $\Delta^-$, for positive and negative sign of the perturbation $J_{\rm{r}}$ respectively. These gaps are found to be independent of the quasiparticle type. 
All results have been obtained up to order $10$ using perturbative continuous unitary transformations method\cite{Knetter00}. For the sake of brevity and readability, we give them in digital form with 16 digits. 
Setting $t=\frac{J_{\rm{r}}}{J_{\rm{p}}}$, one gets

\begin{widetext}
%
%
\begin{align}
	e_0 / J_{\rm{p}} =& \ -1 -0.1075743423260761\, t -0.4800105159959416\cdot 10^{-1} \, t^2 -0.1883684424301069\cdot 10^{-1} \, t^3\\ \nonumber &\  -0.7193204047271116\cdot 10^{-2} \, t^4 -0.3189688657724091\cdot 10^{-2} \, t^5 -0.1825632908475245\cdot 10^{-2} \, t^6\\ \nonumber &\ 
	 -0.1227949920079350\cdot 10^{-2} \, t^7 -0.8687886947035938\cdot 10^{-3} \, t^8 -0.6282297119392671\cdot 10^{-3} \, t^9\\ \nonumber &\ 
	  -0.4676303260709951\cdot 10^{-3} \, t^{10}\ ,\displaybreak[0]\\
 \Delta^+ / J_{\rm{p}} =& \ 1-0.2151486846521522\, t-0.2532895922169015\, t^2-0.1592672177491858\, t^3\\ \nonumber &\ -0.5523232660164432\cdot 10^{-1} \, t^4-0.5488349134970650\cdot 10^{-2} \, t^5-0.7229026627192930\cdot 10^{-2} \, t^6\\ \nonumber &\ -0.2434031543834599\cdot 10^{-1} \, t^7-0.2828066425443511\cdot 10^{-1} \, t^8-0.1659099502072632\cdot 10^{-1} \, t^9\\ \nonumber &\ -0.3692384197073231\cdot 10^{-2} \, t^{10}\ ,\displaybreak[0]\\
 \Delta^- / J_{\rm{p}} =& \ 1+0.2151486846521522\, t+0.8442986407230049\cdot 10^{-1} \, t^2+0.2404995783563480\cdot 10^{-1} \, t^3\\ \nonumber &\ +0.2960780291075518\cdot 10^{-2} \, t^4-0.2662150595365672\cdot 10^{-2} \, t^5-0.3900441571239214\cdot 10^{-2} \, t^6\\ \nonumber &\ -0.4221550724116962\cdot 10^{-2} \, t^7-0.4288071526376318\cdot 10^{-2} \, t^8-0.4168169120295962\cdot 10^{-2} \, t^9\\ \nonumber &\ \ -0.3914732278829173\cdot 10^{-2} \, t^{10}\ .
\end{align}
%
%
\end{widetext}
\mbox{}
\vspace{0.5cm}

\vfill

%

\end{document}